      [1999/12/01 v1.4c Il Nuovo Cimento]
\documentclass{cimento}
\input{psfig.sty}
\def\Om{\mbox{$\Omega_{\rm M}$}}
\def\OL{\mbox{$\Omega_{\Lambda}$}}

\title{Cosmology with Gamma Ray Bursts}
\author{G. Ghisellini\from{ins:oab}\ETC,
G. Ghirlanda\from{ins:oab},
C. Firmani\from{ins:oab}\from{ins:mx},
D. Lazzati\from{ins:col},
        \atque
V. Avila--Reese \from{ins:mx}} 
\instlist{\inst{ins:oab} Osservatorio Astronomico di Brera, 
Via Bianchi 46, I-23807 Merate Italy
  \inst{ins:mx}Instituto de Astronom\'{\i}a, U.N.A.M., A.P. 70-264, 
04510, M\'exico, D.F., M\'exico 
  \inst{ins:col} JILA, Campus Box 440, University of Colorado, 
Boulder, CO 80309--440 USA
}
\PACSes{\PACSit
{98.70.Rz}{Gamma--Ray bursts}
\PACSit{98.80.Es}{Observational cosmology}
}
\begin{document}

\maketitle

\begin{abstract}
Apparently, Gamma--Ray Bursts (GRBs) are all but standard candles.
Their emission is collimated into a cone and the received flux
depends on the cone aperture angle.
Fortunately we can derive the aperture angle through an achromatic
steepening of the lightcurve of the afterglow, and thus
we can measure the ``true" energetics of the prompt emission. 
Ghirlanda et al. (2004) found that this collimation--corrected energy 
correlates tightly with the
frequency at which most of the radiation of the prompt is emitted.
Through this correlation we can infer the burst energy accurately
enough for a cosmological use.
Using the best known 15 GRBs we find very encouraging results
that emphasize the cosmological GRB role.
Probing the universe with high accuracy up to high redshifts, 
GRBs establish a new insight on the cosmic expanding
acceleration history and
accomplish the role of ``missing link" between the Cosmic Microwave
Background and type Ia supernovae,
motivating the most optimistic hopes for what can
be obtained from the bursts detected by SWIFT.
\end{abstract}

\section{Introduction}

The huge emitted power of Gamma Ray Bursts (GRBs) makes them
detectable at any redshift $z<20$.
It is therefore natural to use them as probes of the far universe,
to explore the epoch of reionization 
(see e.g. \cite{ref:barkana}, \cite{ref:lamb})
and to study the properties of the material between them and us, 
through the study of absorption lines in the spectra of their prompt
emission and afterglow.
Furthermore, as it seems very likely, GRBs are intimately connected
with the formation of massive stars, and through GRBs we hope to
shed light on the  star formation history at unprecedented
redshifts, without the limitation of extinction which may plague
samples of distant galaxy resulting from deep surveys (\cite{ref:firmani04},
see also Firmani et al. in these proceedings).
However, until now, the hopes to use them as standard candles
to measure the geometry and kinematics of our universe have
been frustrated by the large dispersion of their energetics,
even when corrected for collimation (\cite{ref:frail}; \cite{ref:bloom}).
This changed with the findings by Ghirlanda et al. (2004a,
hereafter GGL, see also Ghirlanda et al. in these proceedings)
that the spectral properties of GRBs are related with the energy 
radiated within their collimation cones.
This correlation is tight enough to allow the most optimistic
hopes for the cosmological use of GRBs in the SWIFT era.
This ``Ghirlanda" correlation links the energy $E_{\rm peak}$
where most of the prompt radiation is emitted with the
total energy $E_\gamma$ radiated during the prompt phase of a GRB.
Since the emission is collimated into a cone,
$E_\gamma$ is not simply measured by the received fluence.
To find it, one needs to measure the semi--aperture cone angle, 
through the achromatic break in the lightcurve of the afterglow.
The Ghirlanda correlation is very similar to the ``stretching" relation of SN Ia
(\cite{ref:phyllips}): more powerful SN Ia have lightcurves that decay
more slowly.
The Ghirlanda relation is then a sort of Cepheid--like relation:
even if the collimation--corrected energetics of GRBs are
not all equal (i.e. they are not standard candles in the strict sense), 
we have a tool to know each of them (GRBs become ``known candles").

\section{The Hubble diagram}

\begin{figure}
\psfig{figure=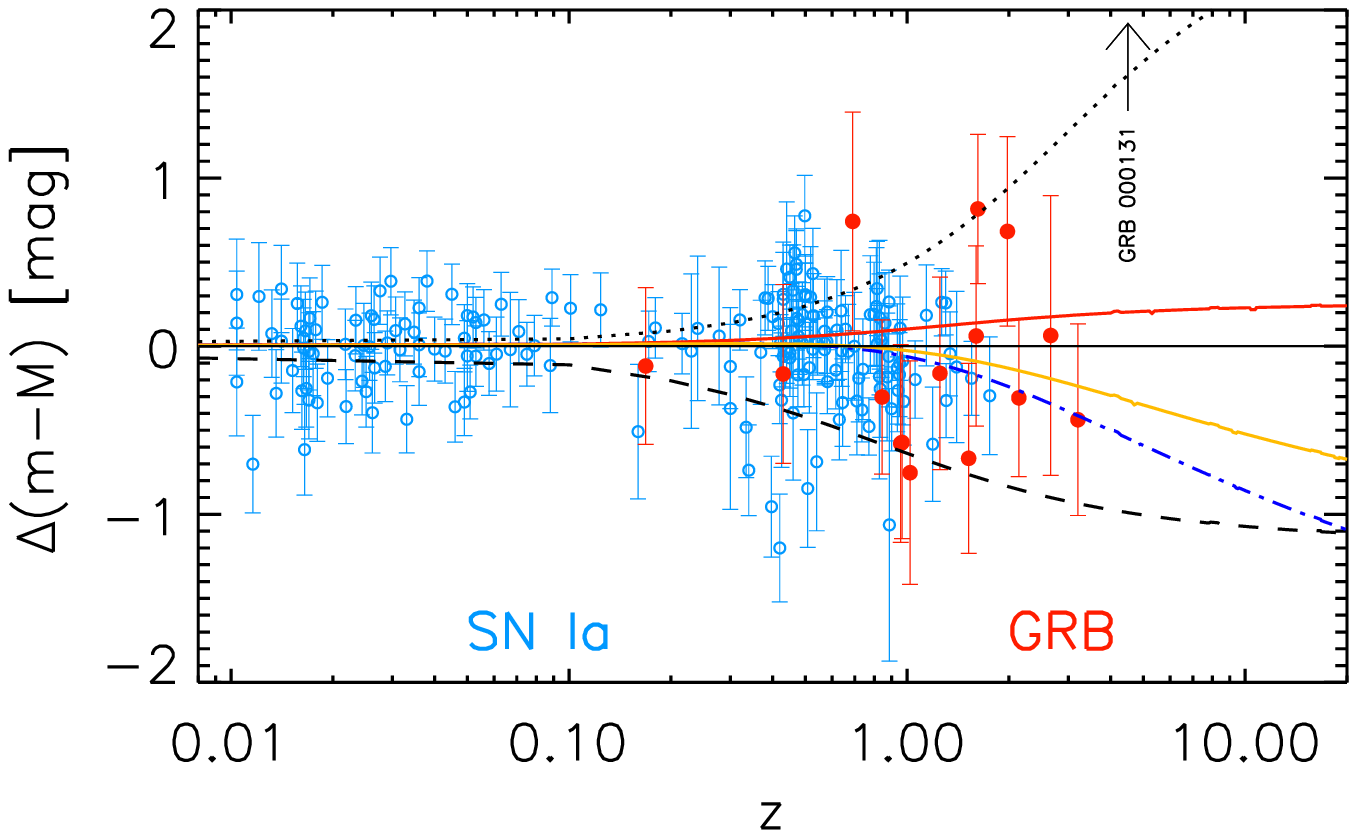,width=14cm,height=8cm}
\vskip -0.5 true cm
\caption{
Residuals of the distance moduli of
SNe Ia and of the 15 GRBs with known redshift and
collimation angle with respect to the 
case \Om=0.27, \OL=0.73.  
Also shown are the differences of various
other cosmological models with respect to the 
\Om=0.27, \OL=0.73 ~case: 
red solid line: \Om=0.2, \OL=0.8; 
orange solid line: \Om=0.37, \OL=0.85; 
blue dot--dashed line: \Om=0.45, \OL=0.95; 
black dashed line: \Om=1, \OL=0; 
black dotted line: \Om=0.01, \OL=0.99. 
The arrow
marks the redshift of GRB 000131 ($z$=4.5) which is the most distant
GRB of known $z$.  From \cite{ref:firmani}.
}
\label{hubble}
\end{figure}

Fig. \ref{hubble} shows the Hubble diagram in the form of residuals with
respect to the specific choice of \Om=0.27 ~and \OL=0.73, for SNe Ia
and GRBs, together with different lines corresponding to different
\Om, \OL ~pairs. 
Since it is likely that GRBs follow the star formation history,
it is also very likely that they exist up to $z\sim$10--20. 
Even at these redshifts, GRBs are easily detectable.
Note that the error bars of GRBs are slightly larger 
than the error bars of SN Ia, but note also that 
the maximum redshifts already measured of GRBs is $z=4.5$,
while SN Ia extends up to a maximum redshift of 1.7.
This is also the maximum foreseen value of the future
SNAP\footnote{http://snap.lbl.gov/} 
satellite, designed to accurately find and observe
distant supernovae to accurately constrain cosmological 
parameters and models.
Needless to say, it is highly desirable to have more than one
class of standard (or known) candles, to cancel possible
evolutionary effects (which are likely different for different
classes of objects) and to cure possible extinction effects
(for GRBs, the hard X--rays of their prompt emission are
virtually unaffected by absorption).
 
The crucial issue about the use of the Ghirlanda relation
for cosmology is what can be called the ``circularity problem":
to find slope and normalization of the correlation,
we are obliged to choose a particular cosmology 
(i.e. a particular pair of \Om, \OL ~values).
The correlation has not (yet) been calibrated with 
low--redshifts GRBs, unaffected by the choise
of \Om ~and \OL, simply because low redshift GBRs
are very few, and we may wait for years (even with SWIFT)
before building a sample numerous enough of low--$z$ GRBs.
In the meantime, we have to deal with the circularity
problem, and in the next section we will describe
the 4 existing methods already suggested.

\section{How to cure the circularity problem}

The Ghirlanda relation has already been used to find constraints
on the cosmological parameters. 
We briefly describe here the four methods that have been proposed and used.
In principle, the cosmological parameters to find are not only
\Om ~and \OL, but also the parameter
$w$ entering in the equation of state of the dark energy
$P=w\rho c^2$ where $P$ and $\rho$ are the pressure and
energy density of the dark energy.
Note that $w$ can be assumed constant  or be  a function of redshift.
One possible parametrization is $w=w_0+w^\prime z$.
The cosmological $\Lambda$ term corresponds to $w_0=-1$ and $w^\prime=0$.
Consider for simplicity this latter case.
To the aim to put constraints on \Om ~and \OL, the basic ideas of the
four methods are the following:

\begin{enumerate}
\item
Dai et al. (2004) assumed that the $E_{\rm peak}$--$E_\gamma$ 
correlation found using a given pair of (\Om, \OL ) values was 
cosmology independent. 
In other words, Dai et al. (2004) assumed that the slope of the 
correlation is fixed (they used $E_\gamma \propto E_{\rm peak}^{3/2}$).
Instead, this slope corresponds to a specific choice 
of \Om, \OL.
This method could be applied if there is a robust theoretical
interpretation of the Ghirlanda relation, putting an heavy weight
to a given slope and normalization of the correlation.

\item 
Ghirlanda et al. (2004b, GGLF hereafter), 
pointing out the circularity problem of the Dai et al. approach,
proposed instead to use the scatter around the fitting line of the
$E_{\rm peak}$--$E_\gamma$ correlation as
an indication of the best cosmology.
In other words, they find a given Ghirlanda correlation using
a pair of \Om, \OL, and measure the scatter of the points
around this correlation through a $\chi^2$ statistics.
Then they change \Om, \OL, find another correlation,
another scatter and another $\chi^2$ which can be compared 
with the previous one.
Iterating for a grid of \Om, \OL ~sets, they can assign to any point
in the \Om, \OL ~plane a value of $\chi^2$, and therefore draw 
the contours level of probability.

\item
GGLF also proposed and used the more classical method of 
the minimum scatter in the Hubble
diagram as a tool to discriminate among different cosmologies.
In practice, this method is almost equivalent to the previous one.

\item 
Firmani et al. (2005) proposed a more advanced method, 
based on a Bayesan approach. 
This is the best method to cure the circularity inherent in the 
use the Ghirlanda correlation.
The basic novelty of this method is that it incorporates the
information that the correlation is unique, even if we do not
know its slope and normalization.
The basic idea is to find the Ghirlanda correlation in a
given point $\bar\Omega$ of the \Om, \OL ~plane, and see how the
entire plane (i.e. any other pair of \Om, \OL) 
``responds" to this correlation: the correlation
remains fixed, but the data points are calculated using the
(running) \Om ~and \OL values. 
The scatter of the points in the Hubble diagram
gives a corresponding $\chi^2$ and we can transform it into a probability.
Then we calculate the correlation in another $\bar\Omega$ point
of the plane, and repeat the procedure.
We repeat that for all (say $N$) $\bar\Omega$ points.
At the end we will have $N$ sets of probabilities.
We can then sum these sets giving a weight to each of them.
The first time we do that we can assign an equal weight.
The sum is a ``probability surface" characterizing each 
$\bar\Omega$ point.
The next time we do this sum, we account for this information
(i.e. there are certain $\bar\Omega$ points more probable than others)
by assigning a weight to each term of the sum equal to
the probability we have derived in the previous cycle.
Then we iterate until convergence, which is reached when the
``probability surface" does not change from one iteration
to the next. An optimization of this method is obtained using a Montecarlo
technique.

\end{enumerate}

\begin{figure}
\vskip -0.5 true cm
\begin{tabular}{ll}
\hskip -1.2 truecm \psfig{figure=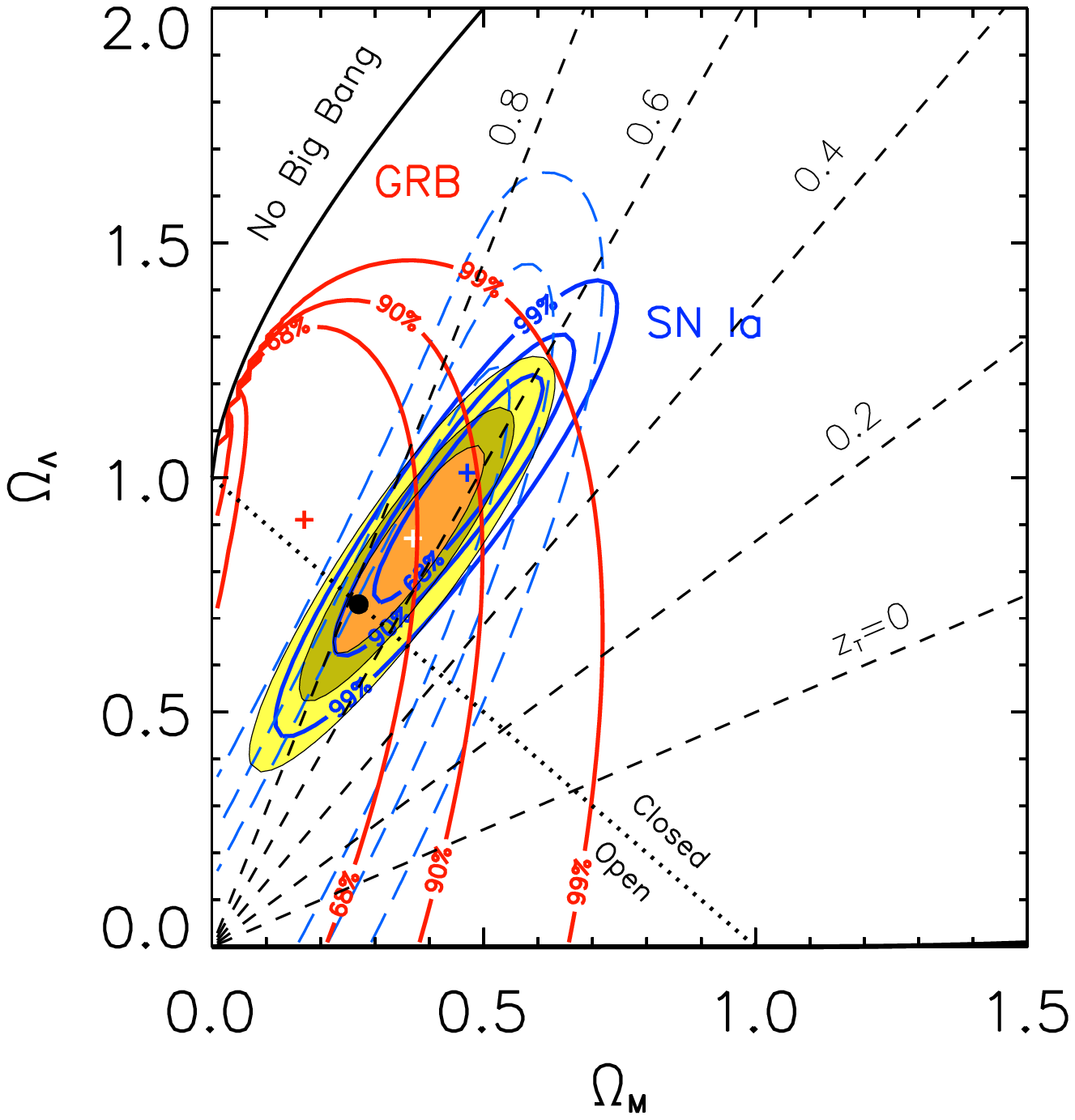,width=8cm,height=9cm} 
& \hskip -1.2 truecm \psfig{figure=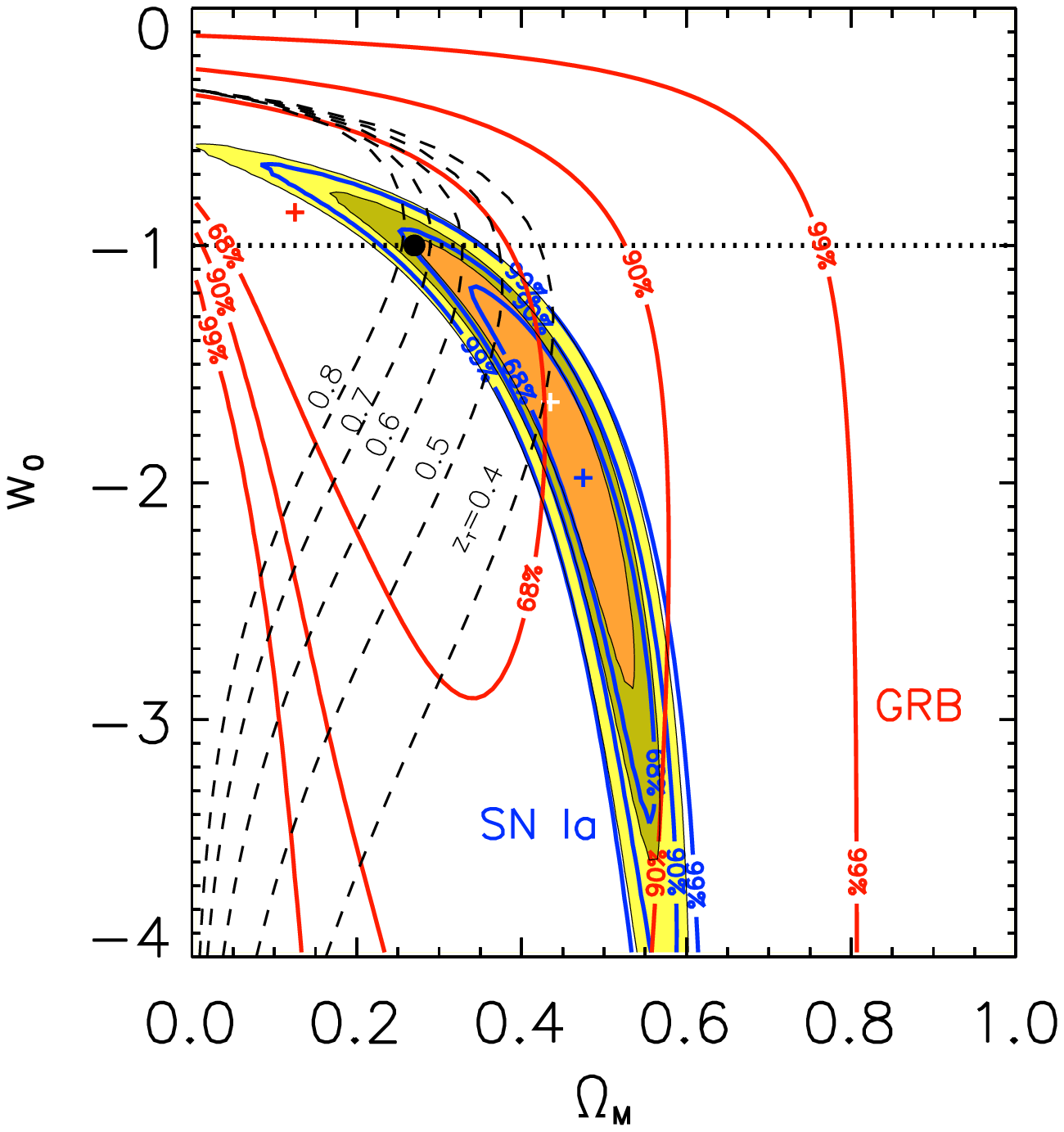,width=8cm,height=9cm} \\
\end{tabular}
\caption{
{\bf Left:}
Constraints in the \Om--\OL ~plane derived from our GRB
sample (15 objects, red contours), from the ``Gold" SN Ia sample (156
objects; blue solid lines, derived assuming a fixed value of $H_0=65$
km s$^{-1}$ Mpc$^{-1}$, making these contours slightly different from
Fig. 8 of \cite{ref:riess04}, and from the subset of SNe Ia at $z>0.9$
(14 objects, blue long dashed lines).  The three colored ellipses are
the confidence regions (orange: 68\%; light green: 90\%; yellow: 99\%)
for the combined fit of type Ia SN+GRB samples.  Dashed lines
correspond to the changing sign of the cosmic acceleration
[i.e. $q(z)=0$] at different redshifts, as labelled.  Crosses are the
centers of the corresponding contours (red: GRBs; blue: SNe Ia, white:
GRB+SN Ia).  The black dot marks the $\Lambda$CDM cosmology.  The
dotted line corresponds to the statefinder parameter $r=1$ 
(see e.g. \cite{ref:sahni}), 
in this case it coincides with the flatness condition. 
{\bf Right:} Constraints on $w_0$ and \Om\ for a
flat cosmology with dark energy whose equation of state is constant  
($w_0=-1$ corresponds to a cosmological constant). 
Colored regions: combined (GRB+SN Ia) constraints (color code as in Fig. 2).
Dashed lines correspond to the changing sign of the cosmic acceleration 
[i.e. $q(z)=0$] at different redshifts, as labelled.  
Crosses are the centers of the SNe Ia contours (blue: SNe Ia
alone, white: GRB+SN Ia). 
The black dot marks the $\Lambda$CDM cosmology.
The dotted line corresponds to the statefinder 
$r=1$. From \cite{ref:firmani}.
}
\label{omega}
\end{figure}

To find constraints on $w_0$ and $w^\prime$ 
the approaches are the same as discussed above,
but bearing in mind that the paucity of data, at present, 
does not allow
to constrain more than two parameters, hence in these
cases one assumes a flat universe and $w^\prime=0$
(and derives constraints on \OL ~and $w_0$), 
or the concordance model (to constrain $w_0$ and $w^\prime$).

\section{Cosmological constraints}

In Fig. \ref{omega} we show the results taken 
from Firmani et al. (2005).
On the left we show the confidence contours on \Om, \OL
~when considering the GRB sample alone (15 objects, red lines),
SN Ia alone (156 objects, blue lines) and the combined
GRB+SN Ia sample (colored regions).
The 68\% confidence contours of the 15 GRBs are less restrictive
than the corresponding SN Ia contours, but note that the 
contours of the combined sample are now consistent with
the concordance model  (\Om=0.3, \OL=0.7) within the 68\% level, 
whereas the SN Ia alone are not.
The dashed blue lines are the contours corresponding to the
14 SN Ia with $z>0.9$.
While the extents of the confidence contour regions are comparable
(the slightly larger errors of GRBs are compensated by their
larger redshifts), the inclination of these ``ellipses" is different, 
and this is due to the different average redshifts of GRBs and this 
subsample of SN Ia, as explained below.
The right panel of Fig. \ref{omega} shows the constraints in the
\Om--$w_0$ plane, once a flat universe is assumed.
Again, note that the GRBs alone do not yet compete with SN Ia,
yet the constraints of the combined sample makes the concordance
model ($w_0=-1$) more consistent, being now fully within 
the 68\% contour levels.

\section{The Cosmic Whirl}
\begin{figure}
\vskip -0.5 true cm
\begin{tabular}{ll}
\hskip -0.5  truecm \psfig{figure=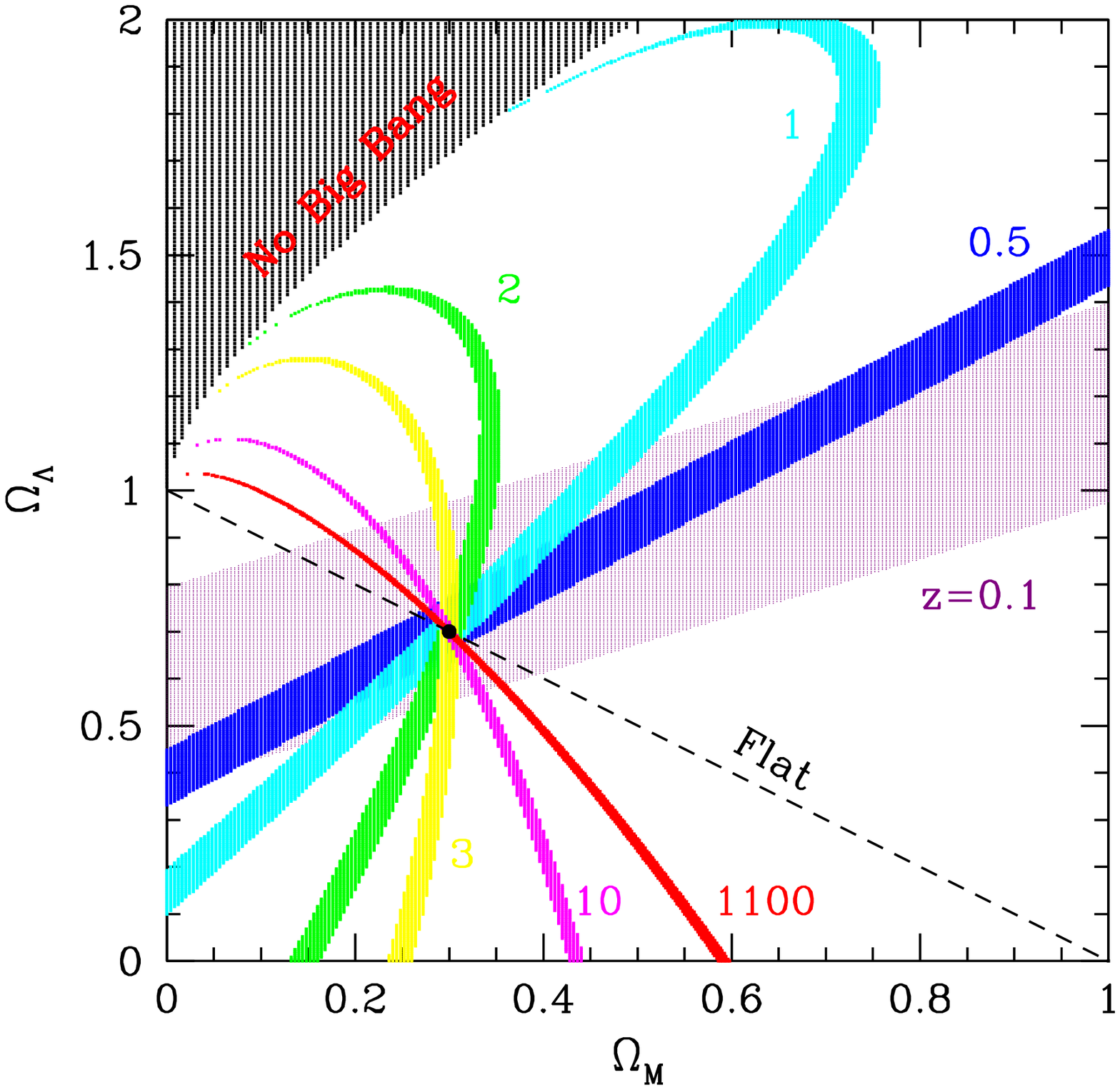,width=8cm,height=9cm} 
& {\hskip -1.7 truecm \psfig{figure=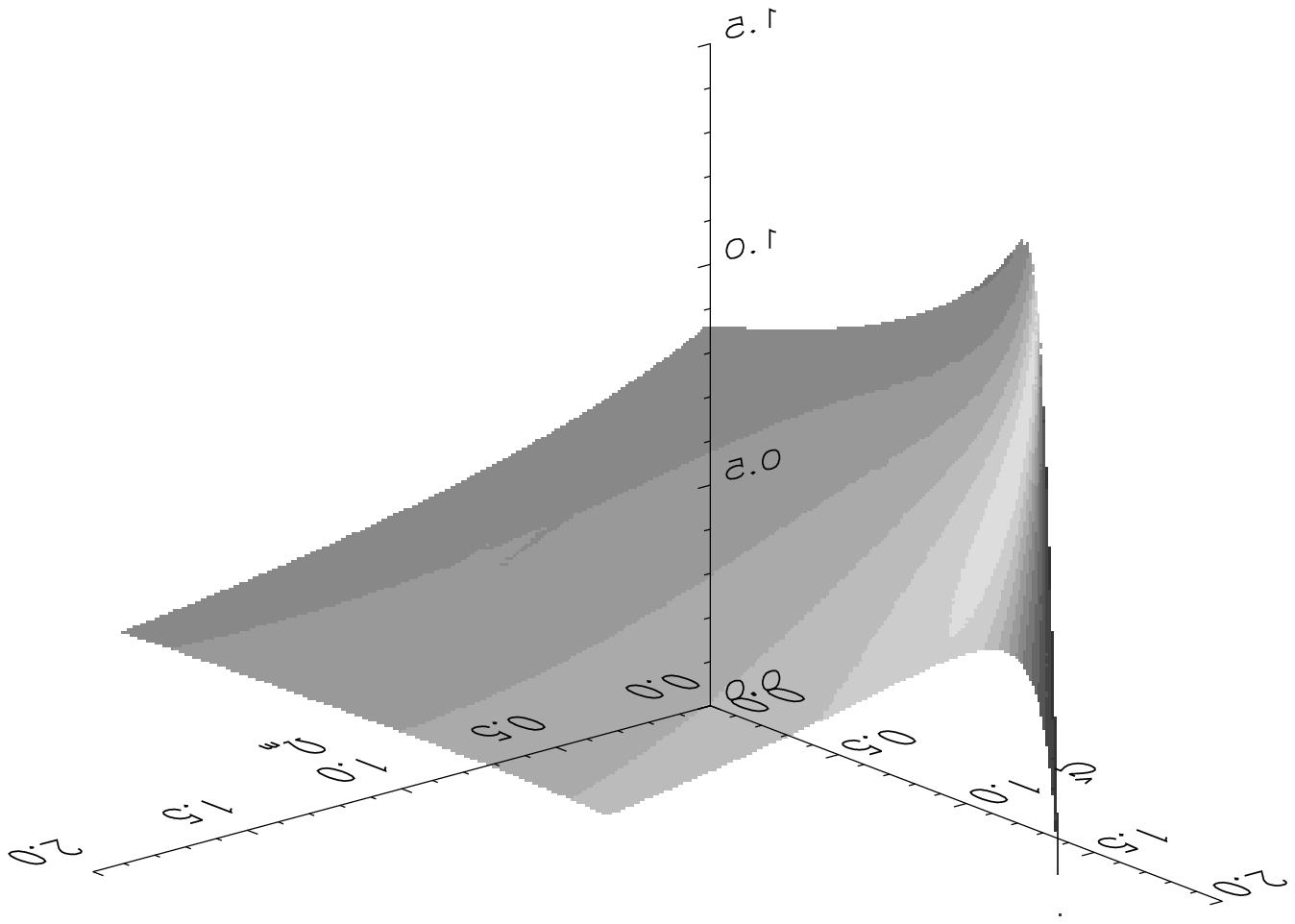,width=8.5cm,height=9cm}} \\
\end{tabular}
%
\vskip -0.6 true cm
\caption{
{\bf Left:}
The Cosmic Whirl. Each stripe represents the loci of points
of luminosity distance differing at most by 1\% to the value
calculated for \Om=0.3, \OL=0.7, for
different values of the redshift, as labelled.
{\bf Right:} The surface of equal luminosity distance calculated for $z=1$.
}
\label{svirgole}
\end{figure}
%
%
%

It is instructive to see the behaviour of the luminosity
distance $d_{\rm L}(z,\Omega_{\rm M}, \Omega_\Lambda)$ 
in the \Om -- \OL ~plane, for different
redshifts.
To this aim, we show in Fig. \ref{svirgole} the ``stripes" 
within which $d_{\rm L}$ changes by 1\% in the \Om -- \OL ~plane, 
for different redshift, assuming  that 
each stripe passes through \Om=0.3 and \OL=0.7. 
Note the following:
\begin{itemize}

\item 
The width of the stripes decreases for larger redshifts.
This is a consequence of the topology of the surfaces
of constant $d_{\rm L}$: at low redshift the surface is 
a gently tilted plane, at high redshifts the surface
is more warped, and there appears a ``mountain" with the peak 
close to \Om$\sim 0.1$ and \OL$\sim 1$.
This is shown in Fig. \ref{svirgole} for $z=1$.

\item 
As a consequence, the ``stripes" at high redshifts are curved,
and at very high redshift they surround the ``mountain peak".

\item
Note that the width of the stripes at high redshifts become
narrower for large \OL --values, as a consequence of the
increasing slope of the $d_{\rm L}$ plane.

\end{itemize}

This example shows that if we have a sample of standard candles 
characterized by a small range in redshifts, the corresponding confidence
contours in the \Om -- \OL ~plane will be elongated in the 
the direction of the stripe of the average redshift of the sample.
This easily explains why the confidence contours derived by using
SN Ia are elongated in the direction of the $d_{\rm L}(z\sim 0.5)$ stripe,
while the contours derived by the WMAP satellite are elongated along
$d_{\rm L}(z\sim 1100)$ stripe.
The contours derived using our GRB sample, of larger average redshift
than SN Ia, are more vertically elongated.
Therefore there is a counterclockwise rotation of the confidence contours
when increasing the average redshift of the standard candles we are using:
only a class of sources spanning a large range of redshifts can give
an accurate value of \Om ~and \OL.

\section{The future}
\begin{figure}
\hskip -1. truecm \psfig{figure=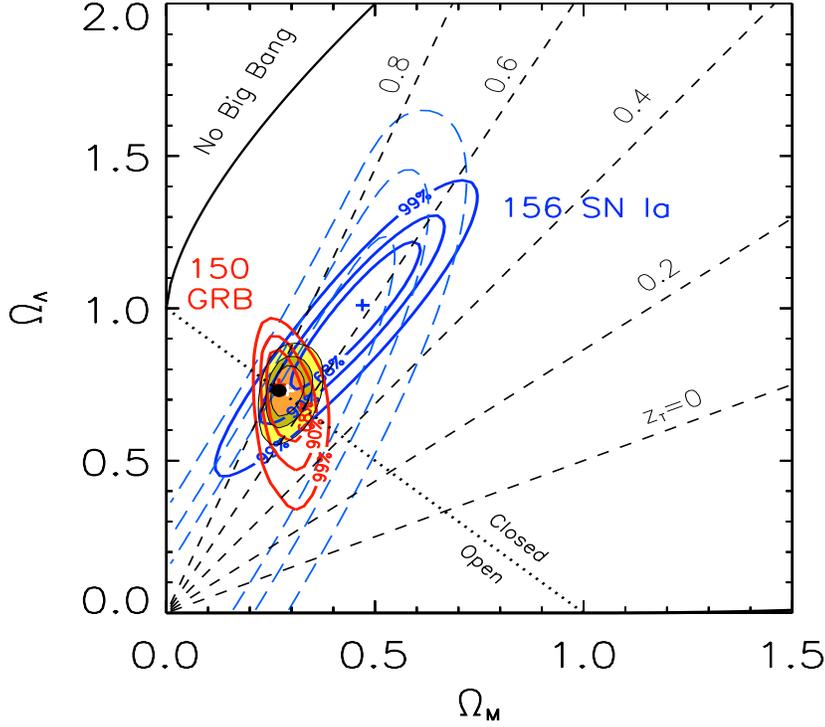,width=13cm,height=11cm} 
\caption{
Contours levels in the \Om, \OL plane for a simulation of 150 GRBs,
according to method 4 (see text and compare with Fig. 2 showing the
results with the current sample of 15 GRBs). 
Blue solid lines are the contours for the 156 SN Ia of
the gold sample of \cite{ref:riess04}, 
blue dashed lines correspond to the subsample of high--$z$ SN Ia,
colored regions correspond to the combined sample of 156 SN Ia and the 
150 GRBs. 
The blue cross is the position of the minimum $\chi^2$ for SN Ia only,
the red and white crosses are the positions of the minimum $\chi^2$ for 
GRB only and the combined GRB+SN Ia sample, respectively.
These crosses are partially hidden by the black dot, marking
the \Om=0.3, \OL=0.7 point.
Black dashed lines correspond to the changing sign of the cosmic acceleration 
[i.e. $q(z)=0$] at different redshifts, as labelled. 
}
\label{simula}
\end{figure}

The constraints posed by WMAP (\cite{ref:spergel}) and SN Ia
(\cite{ref:riess04}, \cite{ref:perlmutter}), together with 
cluster of galaxies (\cite{ref:allen}) have convincingly led 
towards the so called
``concordance cosmological model", a flat universe
with \Om=0.73 ~and \OL=0.27.
So, why bother?
We care because even if we pretend to know how the universe is now,
we do not know how it was. 
This ignorance is one of the obstacles towards the understanding 
of what is the dark energy.
And this is indeed one of the most important and fascinating
challenges of modern (astro)physics.
This is why GRBs can be extremely important, because they
can bridge the gap between relatively close--by supernovae and
the fluctuations of the cosmic microwave background.
They may not be the only tool, since also through the 
integrated Sach--Wolfe effect one hopes to learn how the
universe was in the past. 
But having more than one way to measure the universe is
certainly not redundant, since each way brings its own
uncertainties, selection effects and so on.
Furthermore, we have ahead a few years of SWIFT results,
which will hopefully discover a few hundreds of new bursts
with well measured properties, especially their afterglows
at early times, which will greatly improve the now poor
sample of 15 cosmologically useful GRBs.
The fact that the measured fluences do not strongly correlate
with redshifts (see GGL) means that even for high redshifts GRBs
we can well determine the fluence itself and the spectral parameters
of the prompt emission (i.e. $E_{\rm peak}$).
SWIFT itself will follow the early afterglow with unprecedented
accuracy, and this will certainly help to construct detailed
lightcurves (in X--rays and in the optical) allowing
to pinpoint the jet break time more accurately.
For bursts at $z>5$, therefore in the absence of the optical
lightcurve (the optical flux is strongly absorbed by Lyman $\alpha$ 
absorption) one can derive the jet break time from the X--ray 
and the infrared lightcurves, and in this case robotic IR telescopes
like REM (for the early data, \cite{ref:zerbi}), and larger telescopes (for later
data) will help.
Fig. \ref{simula} shows how the constraints on the cosmological parameters
could improve passing from 15 to 150 bursts, a number comparable 
to the current SN Ia gold sample of \cite{ref:riess04} which will 
be hopefully reached after 2--3 years of SWIFT.
This simulation has been performed (Lazzati et al. in preparation)
assuming a ``luminosity function" of $E_{\rm iso}$ and an appropriate
redshift distribution (see, e.g. \cite{ref:porciani}).
We have  assigned relative errors of
20\% on $t_{\rm break}$, 16\% on $E_{\rm peak}$ and 10\% on 
the fluence (these are the average relative errors for the 15 GRBs).
The average redshift of the simulated sample 
is $\langle z \rangle\sim 2.26$.
This is the reason of the inclination of the ellipses
of the contour levels (cfr. Fig. \ref{svirgole}).

The results shown in Fig. \ref{simula} suggest that GRBs,
reaching larger redshifts than SN Ia, can constrain
\Om, \OL ~better than SN Ia. 
Even tighter constraints can be achieved, of course, 
after a sufficient
number of GRBs are observed at small redshifts,
allowing a cosmology--independent calibration of the
Ghirlanda relation, and/or after a robust theoretical
interpretation is found for this relation.

For an on--line update: www.merate.mi.astro.it/$\sim$ghirla/deep/blink.htm.

\acknowledgments
We thank Annalisa Celotti and Fabrizio Tavecchio for useful discussions 
and the italian MIUR for funding through Cofin  grant 2003020775\_002.

\end{document}